\documentclass[english,prb,reprint,superscriptaddress]{revtex4-1}
\usepackage{letltxmacro}
\LetLtxMacro{\oldauthor}{\author}
\global\long\def\author[#1]#2{\oldauthor{#2}}
\global\long\def\affil[#1]#2{\affiliation{#2}}

\usepackage[bookmarks=true, breaklinks=false, pdfborder={0 0 0},
  backref=false, colorlinks=false]{hyperref}

\usepackage{amsmath}
\usepackage{amssymb}
\usepackage{wasysym}
\usepackage{graphicx}

\begin{document}

\title{Microwave Admittance of Gold-Palladium Nanowires with Proximity-Induced Superconductivity}

\makeatletter
\author[1]{Russell E.~Lake\thanks{russell.lake@aalto.fi}}
\makeatother
\author[1]{Joonas Govenius}
\author[1]{Roope Kokkoniemi}
\author[1]{Kuan Yen Tan}
\author[1]{Matti Partanen}
\affil[1]{QCD Labs, COMP Centre of Excellence, Department of Applied Physics, Aalto University, P.O. Box 13500, FI-00076 Aalto, Finland}
\author[2,3]{Pauli Virtanen}
\affil[2]{Low Temperature Laboratory, Department of Applied Physics, P.O. Box 15100, FI-00076 Aalto University, Finland}
\affil[3]{NEST, Istituto Nanoscienze-CNR and Scuola Normale Superiore, I-56127
Pisa, Italy}
\author[1]{Mikko M\"{o}tt\"{o}nen}
\affil[1]{QCD Labs, COMP Centre of Excellence, Department of Applied Physics, Aalto University, P.O. Box 13500, FI-00076 Aalto, Finland}


\begin{abstract}
We report quantitative electrical admittance measurements of diffusive
superconductor--normal-metal--superconductor (SNS) junctions at
gigahertz frequencies and millikelvin temperatures.
The gold-palladium-based SNS junctions
are arranged into a chain of superconducting
quantum interference devices. The chain is coupled strongly to a
multimode microwave resonator with a mode spacing of approximately
$0.6\mbox{ GHz}$. By measuring the resonance frequencies and quality
factors of the resonator modes, we extract the dissipative and
reactive parts of the admittance of the chain.
We compare the phase and
temperature dependence of the admittance near
$1\mbox{ GHz}$
to theory based on the time-dependent Usadel equations.
This comparison allows us to identify
important discrepancies between theory and experiment that are not
resolved by including inelastic scattering or elastic spin-flip
scattering in the theory.
\end{abstract}

\maketitle

\section{Introduction}
The transport of direct current (dc) between two superconductors (S)
separated by a diffusive normal-metal (N) link is in general well understood both
theoretically and experimentally.\cite{clarke-prl-1968, usadel-prl-1970, likharev-rmp-1979,
  dubos-prb-2001,tero-prb-2002,courtois-prl-2008} At low temperatures
and currents, Andreev reflection \cite{andreev-jetp-1965} leads to the
formation of a gap in the density of quasiparticle states in N and
allows a dissipationless supercurrent to flow.  This gap has been
directly observed in tunnel spectroscopy
measurements.\cite{gueron-prl-1996, lesueur-prb-2008}
Shapiro steps, supercurrent enhancement, and other non-equilibrium effects
under intense microwave irradiation
have also been extensively studied.\cite{clarke-prl-1968, Notarys1973Josephson, Warlaumont1979MicrowaveEnhanced, Lehnert1999Nonequilibrium, dubos-prl-2001, fuechsle-prl-2009, Chiodi2009Evidence, Virtanen2010Theory}

In contrast, the near-equilibrium response of
super\-conductor--normal-metal--super\-conductor (SNS) junctions
to weak microwave radiation
has become an active area of investigation only recently.\cite{galaktionov-prb-2010, virtanen-prb-2011,kos-prb-2013,ferrier-prb-2013,tikhonov-prb-2015}
While the adiabatic contribution to the kinetic inductance
can be calculated from the dc current--phase relation,
at high frequencies both reactive and dissipative
contributions also arise from other mechanisms,
such as driven transitions between quasiparticle states
and oscillation of the Andreev level populations.\cite{virtanen-prb-2011}
Surprisingly however, little experimental data has been published on
the topic thus far.\cite{chodi-sr-2011,dassonneville-prl-2013}
The parameter regimes and materials studied in the
published experiments are very sparse, hence limiting the
extent to which
theoretical predictions\cite{galaktionov-prb-2010, virtanen-prb-2011,kos-prb-2013,ferrier-prb-2013,tikhonov-prb-2015}
can be
tested. In practice, data on the effective inductance and
losses also expedites the process of designing
high-frequency SNS-junction-based circuits, such as the SNS nanobolometer.\cite{govenius-prb-2014,govenius-prl-2016}

Previous experimental studies\cite{chodi-sr-2011,dassonneville-prl-2013}
have probed flux- and temperature-dependent changes in
the linear microwave response
of a superconducting ring with a gold normal-metal inclusion.
The
superconducting ring consisted of ion-beam-deposited tungsten
in the first experiments,\cite{chodi-sr-2011} and sputter-deposited
Nb with a thin Pd layer at the SN interface in the later
experiments.\cite{dassonneville-prl-2013} A single SNS
ring was biased with a dc magnetic flux
and coupled weakly to a multi-mode microwave resonator.
By measuring flux-dependent shifts in the quality factors and resonance
frequencies,
the authors determined how the complex-valued
electrical susceptibility $\chi$ changes.\cite{chodi-sr-2011,dassonneville-prl-2013}
The change in $\chi$, as a function of flux and temperature, was
reported to be in excellent agreement with theoretical predictions
based on Usadel equations and numerical simulations. However, the
implicit offsets in both the real and imaginary parts of the reported
susceptibility prevent a comparison to the theoretically predicted
absolute values of $\mathrm{Re}[\chi]$ and $\mathrm{Im}[\chi]$. They
also prevent the accurate prediction of the effective inductance ($\mathrm{Re}[\chi]^{-1}$)
and the loss tangent ($\mathrm{Im}[\chi] / \mathrm{Re}[\chi]$),
which are the key quantities for any practical high-frequency application of SNS junctions.

In this article, we present
measurements of the SNS junction admittance $Z^{-1}(\omega) = \chi / i \omega$
for gold-palladium based junctions
at angular frequencies $\omega$ of order $2 \pi \times
1\mbox{ GHz}$.  We use a chain of SNS
superconducting quantum interference devices (SQUIDs) with a strong
capacitive coupling to a multimode microwave resonator
with a typical mode spacing of $0.6\mbox{ GHz}$.
Each chain consists of 20 SNS SQUIDs in series.
The strong coupling and the large number of SQUIDs
lead to significant changes in the frequencies and quality factors of
the resonator modes,
which allow
determining $Z^{-1}$ without an offset.
The absence of an offset enables us to show that the Usadel-equation-based
theory we consider cannot simultaneously
explain the observed real and imaginary parts of $Z^{-1}$.

\section{Samples}

\begin{figure*}[t]
\includegraphics[scale=1]{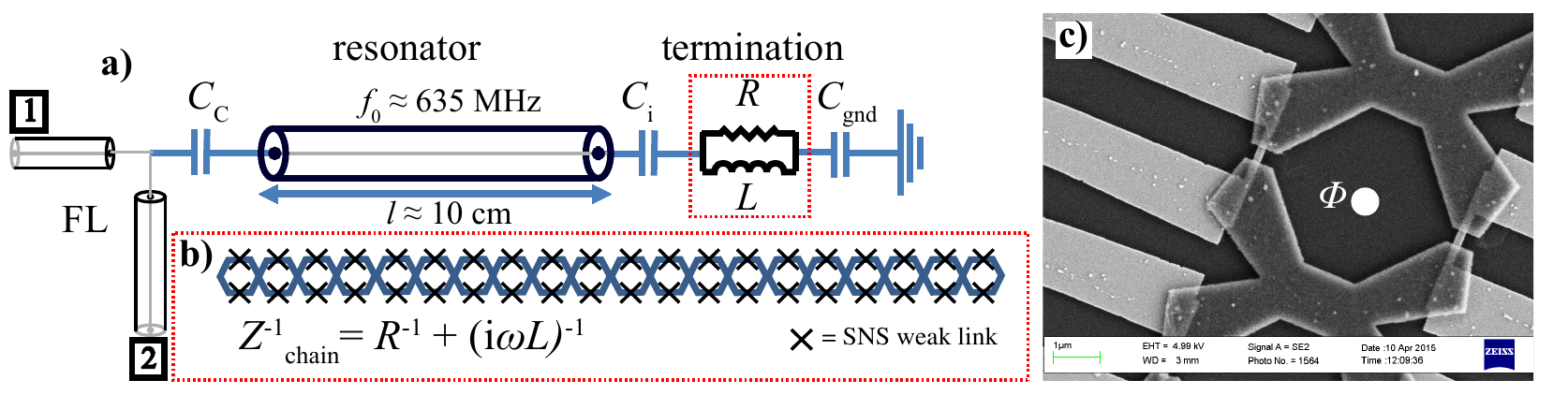}
\caption{
  a) On-chip circuit layout: feedline (FL) between ports 1 and 2,
  transmission line part of the resonator, and termination.
  The resonances are externally damped by the
  capacitive coupling ($C_{\mathrm{C}}$) to the feedline,
  and internally damped by the capacitively coupled
  SQUID chain.
  See Table \ref{sample-table} for further details on parameter values.
  b) Schematic detail of the device under
  test at the termination: 20 SNS SQUIDs in series.
  c) Micrograph of a single SQUID from a sample fabricated
  identically to Sample 2.
  The wide Au-Pd extensions (light colored regions) are heat sinks.
\label{schematic}}
\end{figure*}

We study the linear electrical response of SNS junctions
at millikelvin temperatures in two
samples: Sample 1 and 2.
The gold-palladium nanowires used as the normal-metal are deposited
simultaneously in the same fabrication steps as for our recent
nanobolometer circuits.\cite{govenius-prl-2016} We determine the Au:Pd
atomic ratio of the alloy to be approximately 3:2 using
energy-dispersive X-ray spectroscopy (see Section \ref{exp:nanostructures}).
The nominal junction length $l$ is $300\mbox{ nm}$ and
the nominal cross-sectional area is
$30\mbox{ nm} \times 120 \mbox{ nm}$.
The normal-state resistance of a single junction $R_{\mathrm{N}}=15\ \Omega$ is estimated
based on four-wire dc measurements of reference samples.
From the reference measurements,
we also estimate the upper bound for the contact resistance $R_{\mathrm{B}}$
to be approximately $\apprle 1\ \Omega$.
We only give an upper bound for the contact resistance
because of the uncertainties introduced
by the crossed-wire geometry\cite{Pomeroy2009Negative} of the reference samples.  Reference samples are
similar to those in Figure \ref{schematic}b in Reference \citenum{govenius-prl-2016}. 

The key
parameter determining the strength of the proximity-induced
superconductivity is the Thouless energy $E_{\mathrm{T}} = \hbar D l^{-2}$,
where $D \approx 22\mbox{ cm}^2\mbox{ s}^{-1}$ is the diffusion
constant. In our samples,
the Thouless energy is $E_{\mathrm{T}} \approx k_{\mathrm{B}} \times 190 \mbox{ mK} \approx
h \times 3.9 \mbox{ GHz}$, where $h=2 \pi \hbar$ and $k_{\mathrm{B}}$ is the
Boltzmann constant.
The superconducting sections are 100-nm thick aluminum,
which implies that the energy gap $\Delta$ in the
superconductors is much larger than the Thouless energy ($\Delta/E_{\mathrm{T}}
\approx 13$).  Both S and N parts are fabricated using electron beam
lithography and evaporation.  Further fabrication details are reported
in the Experimental Section \ref{fab}.

The SNS junctions are arranged into a chain of SQUIDs, as shown in
Figure \ref{schematic}.  Each SQUID loop has a relatively small area of $20\ \mu
\mathrm{m}^{2}$ in order to minimize sensitivity to external magnetic
field noise.  In addition, we measured Sample 2 in a double layer
magnetic shield.  The geometric inductance of each
loop is small ($L_{\mathrm{G}} < 10$ pH) compared to the effective inductance of the SQUID, as
verified by the results below.  A dc magnetic flux bias is applied by an external
coil that provides a uniform flux bias $\Phi$ for each SQUID in
the 100 $\mu$m long chain.
Assuming identical junctions,
this phase biases each SNS junction
to $\pi (\Phi - m \Phi_0) / \Phi_0$ at dc,
where $\Phi_0$ is the magnetic flux quantum $h/2e$
and $m$ is the integer that minimizes $|\Phi - m \Phi_0|$.
We note that the flux bias we label as $\Phi = 0$
may be offset from the
true zero flux condition by an integer multiple of $\Phi_0$.

The SQUID chains in the two samples are nominally identical,
except for the addition of heat sinks to Sample 2
(see Figure \ref{schematic}c).
The heat sinks are designed to reduce
the hot-electron effect,\cite{wellstood-prb-1994} i.e.,
the increase of the quasiparticle temperature above
the phonon bath temperature
$T_{\mathrm{b}}$ that we measure.
For each SNS junction, the heat sink
consist of two large ($0.5\ \mu \mathrm{m}^{3}$) reservoirs of gold-palladium
that are
thermally strongly coupled to the junction.

In addition to the SQUID chain, the chip contains a transmission line resonator
(see Figure \ref{schematic}a and Table \ref{sample-table}).
We characterize it by measuring control samples with an open termination,
i.e., samples without the SQUID chain.
From the control measurements, we extract the fundamental
frequency of the transmission line resonator $f_0 \approx 635 \mbox{ MHz}$,
and confirm that the internal quality factor $Q_{\mathrm{i},n} > 10^{4}$
for the resonances we consider ($n \sim 3$).
The latter implies that
we can neglect
the losses in the transmission line
part of the resonator,
and in the
Al$_{2}$O$_{3}$
used as the dielectric material
in the lumped element capacitors ($C_{\mathrm{C}} \mbox{ and } C_{\mathrm{i}}$).
This is valid
because
introducing the SQUID chain lowers $Q_{\mathrm{i},n}$ to the order of $10^2$,
as observed below.
We also deduce the characteristic impedance $Z_0 \approx 39\ \Omega$
of the transmission line from the measured $f_0$, the length of the resonator,
and the design value for the inductance per unit length.

\begin{table*}[]
  \caption{Resonator and coupling capacitor parameters for
  Samples 1 and 2:
  the transmission line resonator length $l_{\mathrm{r}}$
  and the internal (external) load capacitance $C_{\mathrm{i}}$ ($C_{\mathrm{C}}$)
  shown in Figure \ref{schematic}.
  The last column emphasizes that Sample 2 includes additional large
  gold-palladium heat sink reservoirs for enhancing electron--phonon coupling.
\label{sample-table}}
\centering
\begin{tabular}{lllllllll}
\cline{1-1} \cline{2-6} \cline{7-8} 
Sample     &  & $C_{\mathrm{C}}$ (pF) & $C_{\mathrm{i}}$ (pF)      & $l_{\mathrm{r}}$ (mm) & $f_{0}$ (MHz) &$Z_{0}$ ($\Omega$)  & Heat sinks
\\ \cline{1-1} \cline{2-6} \cline{7-8} 
\textbf{1} &  & 0.15         & 14.5                          & 97.2     & 637     & 39  & No                         \\
\textbf{2} &  & 0.44         & 15.2                          & 96.5     & 633     & 39  & Yes         \\ \cline{1-1} \cline{2-6} \cline{7-8} 
\end{tabular}
\end{table*}

\section{Measurement scheme and sample characterization}

We determine the admittance of the
SQUID chain by embedding it as the termination of a long (10 cm)
transmission line microwave
resonator, as illustrated in Figure \ref{schematic}a.
We first determine the resonance frequency $f_n$
and the internal quality factor  $Q_{\mathrm{i},n}$ of each mode $n$
by measuring the frequency-dependent transmission coefficient
${S}_{\mathrm{21}}(f)$ through the feedline.
By comparing $f_n$ and $Q_{\mathrm{i},n}$ to values measured in control samples,
we can determine the admittance of the SQUID chain at multiple frequencies.
Specifically, we use a circuit model (Figure \ref{schematic}a) that
allows extracting the admittance of the SQUID chain $Z_{\mathrm{chain}}^{-1}$
from the response of the combined resonator/SQUID-chain system.
The admittance of each individual SNS junction is then given by
$10 Z_{\mathrm{chain}}^{-1}$,
assuming that the junctions are identical
and that geometric inductance is negligible.

\begin{figure}
\includegraphics{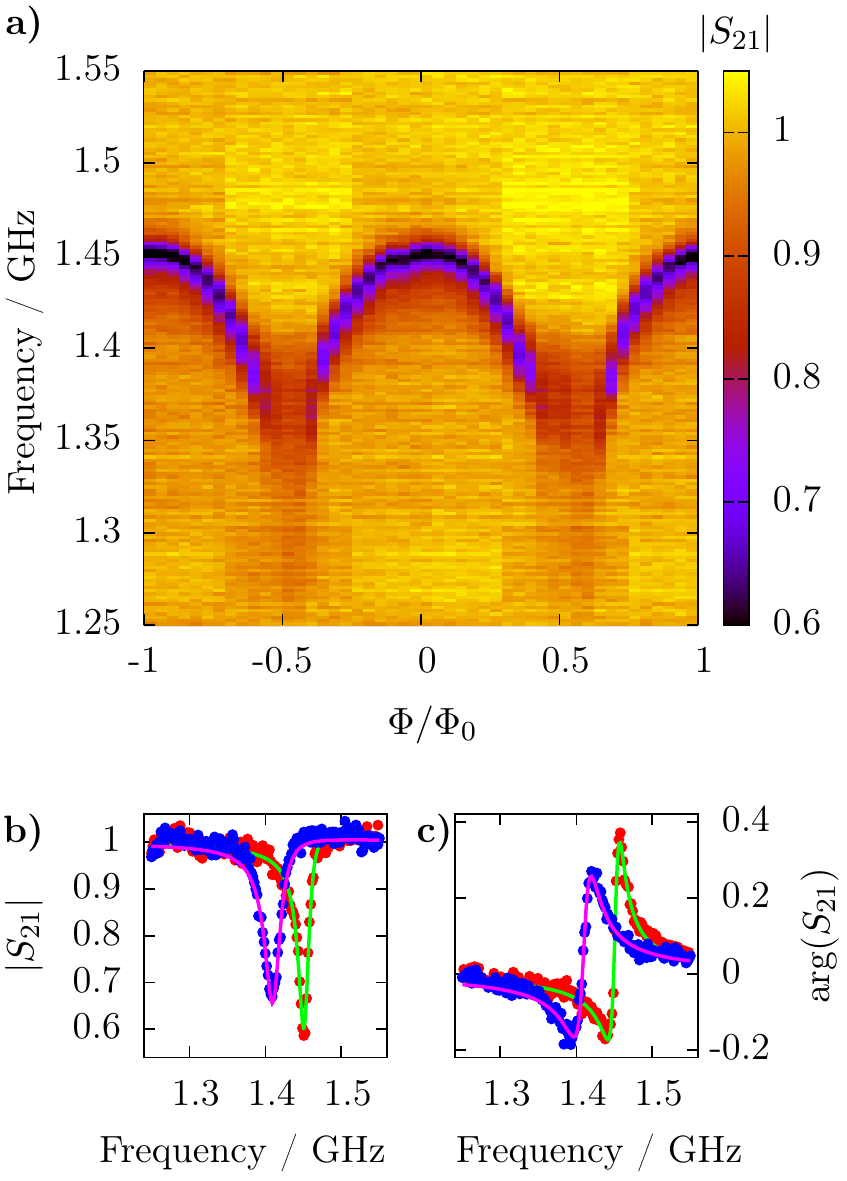}
\caption{a) Magnitude of the normalized transmission coefficient $|{S}_{21}|$
  for the third resonator mode ($n=3$) of Sample 2 at $T_{\mathrm{b}} = 10$ mK as a function of
  frequency and flux-bias $\Phi$.
  b) Traces from a) along $\Phi/\Phi_{0} =
  0$ and $\Phi/\Phi_{0} = 0.3$ with fits (solid lines) to the model in Equation
  \ref{S21}.  c) Same as b) but for the phase $\mathrm{arg}$(${S}_{21}$).
  The extracted resonance frequencies and internal quality factors are
  $f_{3} = (1452.6 \pm 0.4) \mbox{ MHz} $; $Q_{\mathrm{i},3} = 127 \pm 8$ and
  $f_{3} = (1411.1 \pm 0.4) \mbox{ MHz} $; $Q_{\mathrm{i},3} = 78 \pm 4$
  for $\Phi/\Phi_{0} = 0$ and $\Phi/\Phi_{0} =0.3$, respectively.
  The discontinuities in
  the background in a) are artifacts caused by the normalization procedure of ${S}_{21}$
  (see Section \ref{exp:S21norm}).
\label{data1}}
\end{figure}

Figure \ref{data1} shows the normalized transmission through the feedline
at frequencies near $1.4 \mbox{ GHz}$,
probing the third ($n=3$) mode in Sample 2.
The normalization (defined precisely in Section \ref{exp:S21norm})
removes all spurious features in the transmission data that do not
depend on flux.
What remains is the oscillatory flux dependence of the resonance frequency $f_n$
with a period we identify as $\Phi_{0}$.
As the flux bias is increased away from integer multiples of $\Phi_0$,
we measure a decrease in both the resonance frequency
and the loaded quality factor $Q_{0,n}$.
This behavior is more clearly visible in
Figure \ref{data1}b,c with individual slices of transmission data for
$\Phi/\Phi_{0}=0$ and
$\Phi/\Phi_{0}=0.3$.
These changes in the resonance indicate that both the inductance
and the losses increase in the SQUID chain near half-integer values
of $\Phi/\Phi_0$.

\begin{figure}
\includegraphics[width=8.5cm]{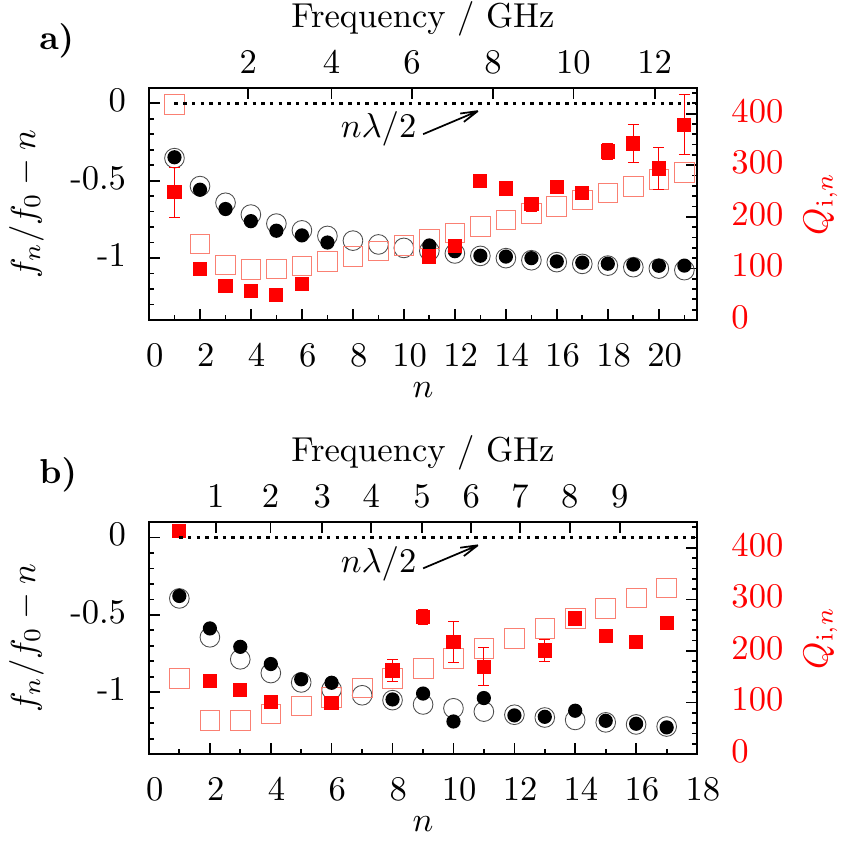}
\caption{
Measured mode shift
$f_n/f_{0}-n$ (filled black circles)
and internal quality factor $Q_{\mathrm{i},n}$ (filled red squares)
for a) Sample 1 and b) Sample 2
at $10\mbox{-mK}$ phonon bath temperature
and zero flux bias.
For reference,
the open markers show the theoretical prediction
based on \mbox{Equation~\ref{circuit-model}}
for a simplistic SQUID chain admittance
$Z_{\mathrm{chain}}^{-1}=R^{-1}+\left(i\omega L\right)^{-1}$
with a constant $R$ equal to
$350\mbox{ \ensuremath{\Omega}}$
($500\mbox{ \ensuremath{\Omega}}$)
and a constant $L$ equal to
$2.5\mbox{ nH}$
($5.0\mbox{ nH}$)
for Sample 1 (2).
The dashed lines emphasize that the mode shift
is zero for all harmonics
of an ideal $\lambda/2$ resonator.
Some $f_{n}$
and $Q_{\mathrm{i},n}$
values could not be experimentally extracted due to
the presence of
nearby parasitic resonances.
\label{modeshift}}
\end{figure}

To extract $f_n$ and $Q_{\mathrm{i},n}$ quantitatively,
we fit the measured normalized transmission for the
$n^{\mathrm{th}}$ mode to the model
\begin{equation}\label{S21}
S_{21}(f) = 1 - \frac{\frac{Q_{0,n}}{Q_{\mathrm{C},n}}-2 i Q_{0,n}
  \frac{\delta f}{f_{n}}}{1 + 2iQ_{0,n} \frac{f - f_{n}}{f_{n}}},
\end{equation}
where $\delta f$ is a fit parameter that characterizes asymmetry,
and the external quality factor $Q_{\mathrm{C},n}$
is governed by the coupling ($C_{\mathrm{C}}$) to the feedline.\cite{geerlings-apl-2012,khalil-jap-2012}
From the obtained fit parameters $Q_{0,n}$ and $Q_{\mathrm{C},n}$, we compute the
contribution of losses due to the SQUID chain as
$Q_{\mathrm{i},n}^{-1}=Q_{0,n}^{-1} - Q_{\mathrm{C},n}^{-1}$.
Figure \ref{modeshift} shows the extracted values of $f_n$ 
and $Q_{\mathrm{i},n}$ for frequencies up to $12 \mbox{ GHz}$ ($n \approx 20$).

In the low-frequency and low-temperature regime,
the SQUID chain behaves like an inductor,
i.e., 
most of the admittance is reactive
($\mathrm{Im}[Z_{\mathrm{chain}}^{-1}] \gg \mathrm{Re}[Z_{\mathrm{chain}}^{-1}]$)
and
$\mathrm{Im}[\omega Z_{\mathrm{chain}}^{-1}]$
varies slowly as a function of the angular frequency $\omega = 2 \pi f$.
Consequently, we parametrize the admittance $Z_{\mathrm{chain}}^{-1}$
as a parallel combination of a resistor and an inductor,
such that $Z_{\mathrm{chain}}^{-1} = R^{-1} + (i \omega L)^{-1}$.
Figure \ref{modeshift} demonstrates that this is a good parametrization
by showing 
qualitative agreement between the experimental data
and predictions for the
mode shifts and quality factors using a simplified model
where $L$ and $R$ are constant.

Let us now discuss the extraction of the admittance $Z_{\mathrm{chain}}^{-1}$
from the measured $f_n$ and $Q_{\mathrm{i},n}$ values.
For an ideal transmission line resonator with open-circuit
conditions at both ends ($C_\mathrm{C} = C_{\mathrm{i}} = 0$),
the $n^{\mathrm{th}}$ mode is located at frequency $nf_{0}$.
In contrast, for the samples with the SQUID chains, the
frequency-dependent reactive, i.e., imaginary  parts of the termination
admittances $i \omega C_{\mathrm{C}}$ and $[(i\omega C_{i})^{-1} +
Z_{\mathrm{chain}}]^{-1}$ lead to the non-zero modeshift
of Figure \ref{modeshift},
that we use to determine $\mathrm{Im}[Z_{\mathrm{chain}}^{-1}]$.
Similarly, the measured $Q_{\mathrm{i},n}$ gives information 
about the dissipative, i.e., real part of $Z^{-1}_{\mathrm{chain}}$. 

Quantitatively, we determine the SQUID chain admittance $Z^{-1}_{\mathrm{chain}}$
from $\omega_n = 2 \pi f_n$ and $Q_{\mathrm{i},n}$ by numerically solving the
trancendental equation
\begin{align} \label{circuit-model}
i\tan & \left[\frac{\omega_n}{2f_{0}} \left(1+\frac{i}{2Q_{\mathrm{i},n}}\right)
              +\arctan\left(Z_{0}\omega_n C_{\mathrm{C}}\right)\right]
 \nonumber \\ &
      =-\frac{Z_{0}}{(i \omega_n C_{\mathrm{i}})^{-1}+Z_{\mathrm{chain}}(\omega_n)},
\end{align}
using the parameters given in Table \ref{sample-table}.
We derive this equation from the circuit model shown in Figure \ref{schematic},
assuming that $Q_{\mathrm{i},n}$ is dominated by losses in the SQUID chain.

Table \ref{losses} shows the $L$ and $R$ extracted
for two examples resonances near 1 GHz.
The reported values provide an important reference for
designing high-frequency devices based on gold-palladium SNS junctions.
That is, they imply that
an effective inductance of a few hundred picohenries per junction
and a loss tangent of a few percent
can be expected around 1 GHz at millikelvin temperatures.

\begin{table}[]
\centering
\caption{SQUID chain admittance $R^{-1} + (i \omega L)^{-1}$
and corresponding loss tangent $\omega L/R$
measured at $T_{\mathrm{b}} = 10$ mK
and $\Phi = 0$ for the second (third) resonance
in Sample 1 (2).
The effective inductance (resistance)
per single SNS junction
is $L/10$ ($R/10$).}
\begin{tabular}{lllllll} 
\cline{1-2} \cline{3-5}\cline{6-7}
 Sample & $f_n$ (GHz) & & $R$ ($\Omega$)                & $L$ (nH)&             & $2 \pi f_n L / R$ \\ \cline{1-2} \cline{3-5}\cline{6-7}
\textbf{1}      & 0.914          &    &        $310 \pm 30$          & $3.4 \pm 0.3$ & & $0.062$       \\
\textbf{2}      & 1.452        &     &        $590 \pm 50$           & $3.1 \pm 0.7$ & & $0.048$      \\\cline{1-2} \cline{3-5}\cline{6-7}
\end{tabular}

\label{losses}
\end{table}

\section{Theory}

In the next section, we compare the experimental results to theoretical predictions \cite{virtanen-prb-2011}
based on the
time-dependent Usadel equation.\cite{usadel-prl-1970} In the
low-frequency and low-temperature regime $\hbar \omega,k_{\mathrm{B}}T \apprle E_{\mathrm{T}}$
considered below,
the imaginary part of the admittance of the
junction is expected to be mostly determined by the adiabatic Josephson
inductance associated with the supercurrent,
i.e., the $\Phi$ derivative of the dc supercurrent.
The real part, on the other hand,
mainly arises from driven quasiparticle transitions in the
junction.
The availability of such transitions is sensitive to the density of quasiparticle states.
In particular, the presence of a proximity-induced energy gap
$E_{\mathrm{g}} \sim E_{\mathrm{T}}$ in the density of states should lead to an
exponential increase in the resistance
as $k_{\mathrm{B}}T$ decreases below $E_{\mathrm{g}}$.

However, the low-temperature values of $L$ and $R^{-1}$
we measure (Table \ref{losses})
are dramatically larger than those predicted
using the parameters considered in Reference \citenum{virtanen-prb-2011}.
This is evident from a cursory comparison of Figure 1 in Reference \citenum{virtanen-prb-2011}
to our $(\omega L/10)^{-1} \sim 6 R_{\mathrm{N}}^{-1}$ and $(R/10)^{-1} \sim 0.3 R_{\mathrm{N}}^{-1}$.
The inductance per junction $L/10 \sim 300 \mbox{ pH}$ is also
an order of magnitude higher than the expected adiabatic Josephson inductance
$L_{\mathrm{J}} = [2\partial_\Phi I_{\mathrm{s}}(\Phi)]^{-1} \sim 50 \mbox{ pH}$,
where we approximate the dc supercurrent
$I_{\mathrm{s}}(\Phi)$ as $I_{\mathrm{c}} \sin(\pi \Phi/\Phi_0)$
and the critical current $I_{\mathrm{c}}$ as the ideal value $6.7 E_{\mathrm{T}} / e R_{\mathrm{N}}$
for $\Delta/E_{\mathrm{T}} = 13$. \cite{dubos-prb-2001}
Moreover---in the results below---we observe a weak temperature dependence of $R(\Phi = 0)$ measured near 1 GHz, 
which is in stark contrast to the theoretically predicted exponential dependence.

The observed values of $L$ and $R^{-1}$ imply that the
the proximity-induced superconductivity is significantly
weaker than expected.
We consider two distinct scattering mechanisms as potential
explanations for this.
First, we 
include dephasing due to inelastic scattering
by choosing a
phenomenological relaxation rate $\Gamma$.\cite{virtanen-prb-2011} 
Second, we include a spin-flip scattering rate $\Gamma_{\mathrm{sf}}$,
which could arise
from dilute magnetic impurities in the weak link.\cite{abrikosov-jetp-1960}
Specifically, we include the spin-flip scattering 
as an additional self-energy
$\check{\sigma}=-\frac{i}{2}\hbar\Gamma_{\mathrm{sf}}\hat{\tau}_3\check{g}\hat{\tau}_3$
in the equations defined in Reference \citenum{virtanen-prb-2011}.
Although quantitative details differ,
both of the scattering mechanisms generally lead to increased dissipation and increased inductance. Increased dissipation occurs mainly due to the suppression of $E_{\mathrm{g}}$, while increased inductance occurs mainly due to the increase in $L_{\mathrm{J}}$.

Theoretical work on the microscopic origin of
the scattering rates in disordered
metals is reviewed in
References \citenum{giazotto-rmp-2006} and \citenum{rammer-rmp-1986}.
Experiments have also been performed with high-purity metal
wires. \cite{pierre-prb-2003,huard-ssc-2004} However, we are not aware
of measurements on the gold-palladium alloy used
here, which prevents direct comparison to existing literature.
Instead, our goal is to estimate the scattering rates required
for a qualitative match to the experimental results.
We find that in order to reproduce the
experimentally observed $L$ or $R$,
the phenomenological rates
$\Gamma$ and $\Gamma_{\mathrm{sf}}$ must be large, i.e., 
comparable to $E_{\mathrm{T}}/\hbar$ and $k_{\mathrm{B}}T/\hbar$.

Inelastic scattering and spin-flip scattering
are not the only possible explanations for observing proximity-induced
superconductivity that is weaker than what is predicted
by the ideal Usadel-equation-based theory.
While we do not attempt to exhaustively cover all candidates,
we note that the SN contact resistance in our samples
is much smaller ($R_{\mathrm{B}} \apprle 1\ \Omega$)
than the normal-state resistance ($R_{\mathrm{N}}=15\ \Omega$).
While the smallness of the ratio $R_{\mathrm{B}}/R_{\mathrm{N}}$
does not conclusively exclude explanations based on
imperfect interfaces,
it limits them significantly. \cite{tero-prb-2002,hammer-prb-2007}

\section{Temperature and flux dependence near 1 GHz}
\label{results}

Below, we compare the predicted and observed
dependences of $Z_{\mathrm{chain}}$ on the bath temperature and magnetic flux.
We choose to analyze two
low-$n$ resonances near 1 GHz,
mainly because the $L$ values we extract for them
suffer the least from the uncertainty in $f_{0}$.

\begin{figure}
\includegraphics[width=8.5cm]{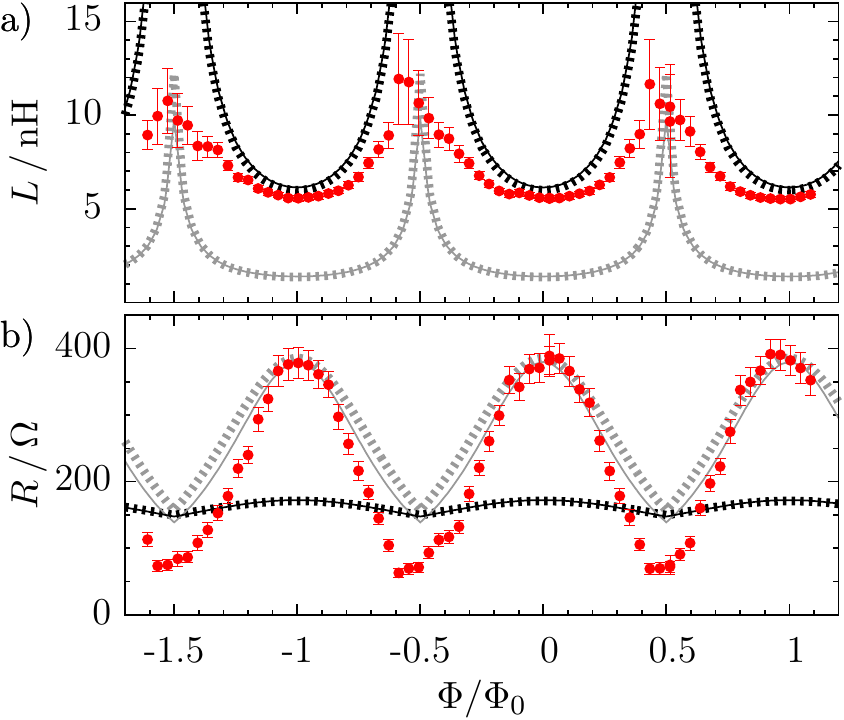}
\caption{Flux-dependent a) $L$ and b) $R$ measured near $1.4 \mbox{
    GHz}$ of the SQUID chain in Sample 2 at 195 mK.  Dashed lines are theoretical
  calculations
  that include only inelastic scattering corresponding to a very high
  scattering rate $\Gamma = 8 k_{\mathrm{B}}T_{\mathrm{b}}/\hbar$
  (black dashed line) and a moderately high scattering rate $\Gamma
  = 2.3 k_{\mathrm{B}}T_{\mathrm{b}}/\hbar$ (gray dashed line).  Solid lines are
  calculations that include very strong elastic spin-flip scattering,
  $\Gamma_{\mathrm{sf}} = 15 E_{\mathrm{T}} / \hbar$ (black solid line), and moderately strong
  scattering, $\Gamma_{\mathrm{sf}} = 4.5 E_{\mathrm{T}} / \hbar$ (gray solid
  line).  Both solid lines include an additional low inelastic
  rate of $\Gamma = 0.05 k_{\mathrm{B}}T_{\mathrm{b}}/\hbar$.
\label{varphi}}
\end{figure}

Figure \ref{varphi} shows the measured flux dependence of $R$ and $L$
for the third ($n=3$) resonance in Sample 2.
The bath temperature is $T_{\mathrm{b}} = 195$ mK, which should be
high enough for neglecting the hot-electron effect, i.e.,
for assuming that $T \approx T_{\mathrm{b}}$.
As expected, we observe that $R$ and $L$ are periodic in flux,
and that the inductance $L$ and the loss tangent $\omega L/R$ are
minimized (maximized) at integer (half-integer) values of
$\Phi/\Phi_0$.

Figure \ref{varphi} also includes theoretical
predictions for two different rates of inelastic scattering.
The weaker of the two
rates ($\Gamma = 2.3 k_{\mathrm{B}}T_{\mathrm{b}}/\hbar$)
reproduces $R(\Phi=0)$ well and
gives a reasonable prediction for
its flux-dependent oscillations.
Furthermore, if we could only measure changes in $L$,
we might conclude that the predicted flux modulation of $L$
is in fair agreement with the experimental data
for this moderate value of $\Gamma$.  However, the
absolute value of the prediction for $L(\Phi)$ is several times
smaller than the observed value
at nearly all flux values.
This highlights the importance of
measuring $L$ and $R$ without offsets if theories are to be
rigorously tested.  Note that we can improve the agreement between the
predicted and measured $L(\Phi)$, especially around integer values of
$\Phi/\Phi_0$, by using a very strong inelastic scattering rate of
$\Gamma = 8 k_{\mathrm{B}}T_{\mathrm{b}}/h$ in the theoretical
calculation.  However, this value of $\Gamma$ leads
to a clear disagreement in
the amplitude of the oscillations in 
$R(\Phi)$ as shown in Figure \ref{varphi}b.

Figure \ref{varphi} also shows the theoretical predictions
that include
strong spin-flip scattering.
By choosing $\Gamma_{\mathrm{sf}}$ appropriately,
the predictions become nearly identical to
the case of strong inelastic scattering.
Therefore, the conclusions of the previous paragraph also apply
to predictions where scattering is spin-flip dominated.
Furthermore, the similarity of the predictions shows that,
in this parameter regime, the source of additional dephasing
is unimportant.

\begin{figure}
\includegraphics[width=8.5cm]{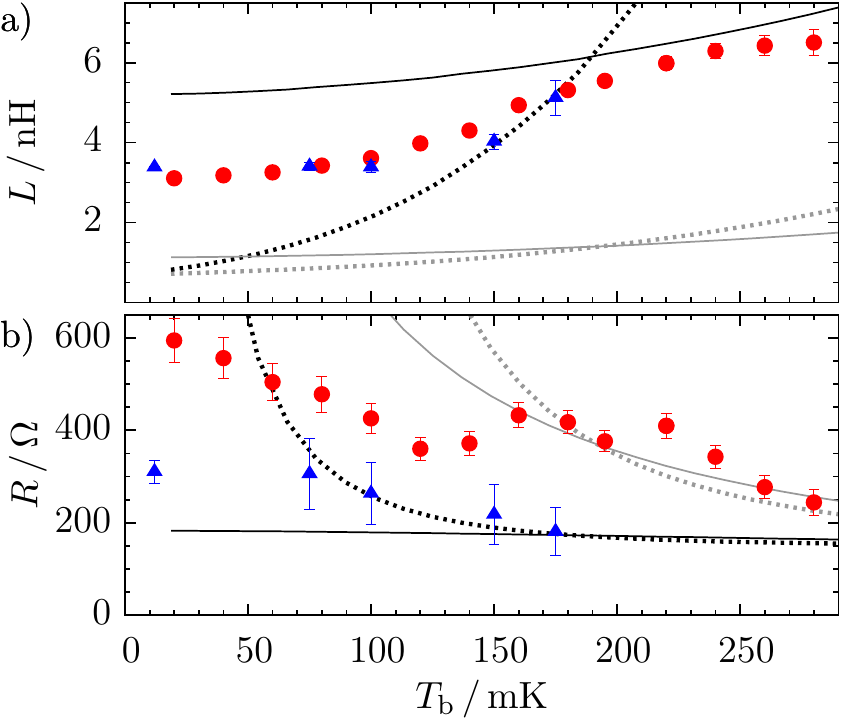}
\caption{Temperature dependence of a) $L(\Phi=0)$ and b) $R(\Phi=0)$
  for Sample 1 near 0.9 GHz (filled triangles) and for Sample 2 near
  1.4 GHz (filled circles).
  The scattering rates for the theoretical predictions
  are given in Figure \ref{varphi}.
\label{varT}}
\end{figure}

To gain further insight,
we study the temperature dependence of $R(\Phi=0)$ and
$L(\Phi=0)$ for one resonance from each sample near 1 GHz (see Figure \ref{varT}).
In addition to the measured data points,
Figure \ref{varT} shows theoretical predictions
with scattering parameters that---at 195 mK---are
identical to those in Figure \ref{varphi}.
However, we note that considerable freedom remains in choosing the
temperature dependence of the scattering rates.
Rigorously justifying a particular temperature scaling
would require knowledge of the specific microscopic mechanism
responsible for the scattering.
However, as the theoretical predictions
already disagree with the measured results
at the phenomenological level
at a fixed temperature (Figure \ref{varphi}),
identifying any specific microscopic mechanism seems implausible.
As instructive examples,
we choose $\Gamma \propto T$
and a constant $\Gamma_{\mathrm{sf}}$ in Figure \ref{varT}.
Unsurprisingly, none of the predictions simultaneously matches
the observed temperature dependence of $L$ and $R$.
Nevertheless, the experimental data in Figure \ref{varT}
may serve an important role in testing alternative theories
in the future.

\section{Conclusions}
The main discrepancy between theory and experiment can be summarized
as follows.  The proximity effect at $T \apprle E_{\mathrm{T}} / k_{\mathrm{B}}$ and $\omega
\sim E_{\mathrm{T}} / \hbar$ is weaker than what is predicted by theory based
on the Usadel equation.\cite{virtanen-prb-2011} This disagreement
manifests itself experimentally as measured $R$ values that fall below
theoretical predictions and measured $L$ values that exceed
theoretical predictions.
As potential candidates for such loss of coherence, we
considered inelastic scattering and spin-flip scattering in the weak link.
However, we did not find choices of $\Gamma$ or
$\Gamma_{\mathrm{sf}}$ that would provide simultaneous agreement in
$L$ and $R$, neither in terms of flux dependence at a fixed bath
temperature, nor in terms of temperature-dependence at zero flux bias.
Furthermore, the scattering rates required for a match
in either $L$ or $R$ are larger than expected for, e.g.,
electron--electron scattering in disordered systems.\cite{rammer-rmp-1986}

We note that the discrepancies shown here are not in direct contradiction with the
previous experiments\cite{chodi-sr-2011,dassonneville-prl-2013} and
that both the SNS junctions and the measurement scheme presented here
are very different from these preceding studies.
Firstly, the weak link material is different than
in the previous experiments.
We cannot rule out the possibility of
effects specific to gold-palladium \cite{mcginnis-prb-1985}
that reduce coherence in
the weak link.
Secondly, we measure
both the reactive and dissipative components of the electrical admittance
without arbitrary offsets.
In contrast, only changes in the admittance have been previously reported.
Thus, our experimental technique provides a more
stringent test of the accuracy of the theory and reveals
quantitative disagreements more easily.

In conclusion, we reported measurements of microwave frequency admittance
for gold-palladium SNS junctions,
together with
a comparison to quasiclassical theory for diffusive SNS weak links.
These discrepancies between measurement results and theoretical predictions
suggest that dephasing caused by inelastic scattering,
or elastic spin-flip scattering,
is probably not the correct mechanism
for explaining why
the proximity-induced superconductivity
is weaker than expected in our gold-palladium SNS junctions.
Further theoretical work is required for reaching simultaneous agreement for the magnitude,
temperature dependence, and flux dependence
of both the dissipative and reactive parts of the admittance.
Mechanisms that may need to be taken into account
include imperfect interfaces \cite{tero-prb-2002, hammer-prb-2007},
electron--electron and fluctuation effects
in low-dimensional superconducting structures, \cite{Narozhny1999Theory, Semenov2012Dephasing}
and paramagnon interaction. \cite{kontos-prl-2004}
Magnetic effects could be particularly important in SNS junctions
that include palladium,
which is paramagnetic in bulk and
can even become ferromagnetic in nanoscale particles.
\cite{Shinohara2003Surface, Sampedro2003Ferromagnetism}
In general, the
relationship between microscopic materials properties and coherence at
microwave frequency in normal-metal Josephson junctions
should be clarified, both experimentally and theoretically.
A productive experimental
approach may be to first investigate systems such as Nb/Cu weak links that,
based on previous dc experiments,\cite{dubos-prb-2001,jabdaraghi-apl-2016}
are expected to behave in an ideal fashion at dc.

\section{Experimental Section}\label{exp}

\subsection{Device fabrication}\label{fab}

\subsubsection{Resonators}
The substrates are 4'' (0.5-mm thick) high-resistivity (${>} 10^{4}\ \Omega \mathrm{cm}$) Si wafers
with 300 nm of thermal oxide.  First, a niobium thin film (thickness
200 nm) is sputter deposited on the entire wafer.  Next, the coplanar
waveguide (CPW) structures are defined with AZ5214E positive
photoresist that is reflowed at $150\ ^\circ\mathrm{C}$ for 1 min to ensure a
positive etch profile of the resulting Nb features. Then CPWs are etched
with an rf-generated plasma under a constant flow of SF$_{6}$(40
sccm)/O$_{2}$(20 sccm) gases at constant
power.\cite{curtis-jvsta-1993} The remaining resist is removed with
solvents and an additional O$_2$-plasma cleaning step.  The 4'' wafer
is then coated with a protective layer of resist and pre-diced with
partial cuts along device pixel outlines on the back of the wafer.

\subsubsection{Capacitor dielectric}
The Al$_{2}$O$_{3}$ dielectric
for the on-chip Nb-Al$_{2}$O$_{3}$-Al capacitors
$C_{\mathrm{C}}$, $C_{\mathrm{i}}$, and $C_{\mathrm{gnd}}$
is formed by atomic
layer deposition with 455 cycles in a H$_{2}$O/TMA process at 200
$^{\circ}$C resulting in a thickness of 42 nm.
The thickness was verified in ellipsometry using
an index of refraction $n_{\mathrm{Al}_{2}\mathrm{O}_{3}} = 1.64$.
Measurements of reference Nb-Al$_{2}$O$_{3}$-Al capacitors
yield a capacitance per unit area of
$1.4\ \mathrm{fF} \mu\mathrm{m}^{-2}$.

\subsubsection{Nanostructures}
\label{exp:nanostructures}
The gold-palladium nanowires and aluminum superconducting leads are
fabricated by electron beam lithography in two separate
evaporation/liftoff steps.
In the first
step, gold and palladium pellets are evaporated from the same crucible
with an electron beam heater.
Afterward, unwanted Au-Pd is lifted off with organic solvents.
Prior to the evaporation of the Al leads, samples are cleaned
in situ with an Ar sputter gun.  Finally, after liftoff of the Al film,
individual resonator pixels are snapped along the pre-diced lines and
packaged for measurement.

\begin{figure}[!hbtp]
\includegraphics[width=8.5cm]{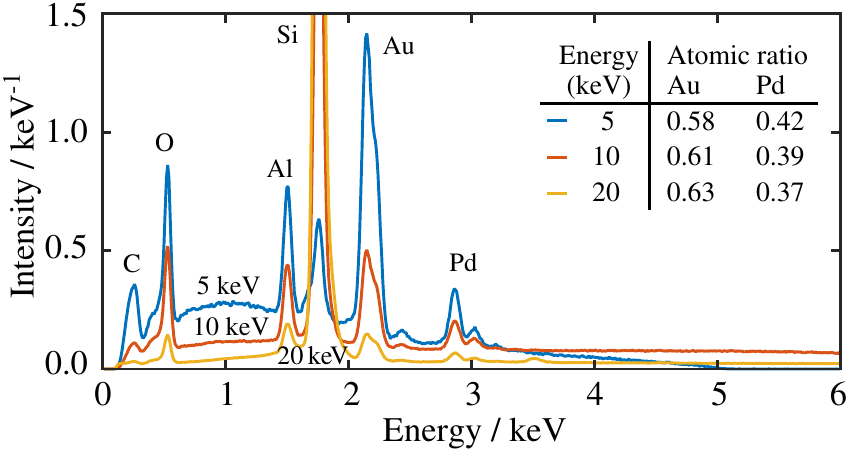}
\caption{ Measured composition of a section of gold-palladium alloy
  evaporated together with Sample 2. The material stack from top to
  bottom is gold-palladium (30 nm), $\mathrm{Al}_2\mathrm{O}_3$ (42
  nm), $\mathrm{Si}\mathrm{O}_{2}$ (300 nm), and Si (500 $\mu$m).
  For clarity, we crop
  the strong signal for Si at high electron beam energies.
  It peaks at 2.8 $\mathrm{keV}^{-1}$ (5.4 $\mathrm{keV}^{-1}$) for 10 keV (20 keV).
\label{AuPd}}
\end{figure}

The chemical composition of the gold-palladium material is determined with
energy-dispersive X-ray spectroscopy for incident electron beam
energies 5 keV, 10 keV, and 20 keV (Figure \ref{AuPd}). The average
Au:Pd atomic ratio (weight ratio) is approximately 3:2 (3:1)

\subsection{Cryogenic measurements}
Measurements are carried out in a commercial cryostat with the base
temperature of 10 mK.  The transmission coefficient was probed with a vector network
analyzer. The device input line had $>$100 dB fixed attenuation.
For all measurements, the output signal is
amplified by a broadband low-noise cryogenic amplifier and by
additional room temperature amplifiers.
For some measurements (e.g.~Sample 2, $n=3$) two cryogenic isolators
are placed on the base cooling stage
between the low-noise cryogenic amplifier and the sample.
Each sample was placed in a
custom printed circuit board and sealed within a metal enclosure.  The
external flux coil consists of a superconducting solenoid with
100 turns that is fixed outside the metal
enclosure. One $\Phi_{0}$ period in Figure \ref{varphi} and \ref{varT} corresponds to a
current change of $\Delta I_{\mathrm{mag}} \approx 7$ mA through the
coil. Magnetic shielding surrounds both the enclosure and flux-coil in the case of Sample 2.

The measurement power incident at the
transmission line input is approximately -128 dBm
for the data shown in Figure \ref{varphi} and \ref{varT}.  This drives a
current of roughly 5 nA through the SQUID chain at $\Phi=0$ for $n=3$
of Sample 2.
This is far below the estimated critical current of the SQUID chain.
Furthermore, experimentally we ensure that we measure the linear response
by making sure that the measured $S_{21}$ is not sensitive to factor-of-two
changes in the measurement power.

\subsection{Normalized transmission coefficient}
\label{exp:S21norm}
We define the normalized transmission coefficient ${S}_{21}$ as
${S}^{\prime}_{21}(\Phi) / [ {S}^{\prime}_{21}(\Phi_{\mathrm{ref}}) / {S}_{21,\mathrm{fit}}(\Phi_{\mathrm{ref}}) ]$,
where ${S}^{\prime}_{21}(\Phi)$ is the full transmission coefficient,
including contributions from the cabling and other external circuitry,
and ${S}_{21,\mathrm{fit}}$ is its best fit to Equation \ref{S21}.
Dividing by ${S}^{\prime}_{21}(\Phi_{\mathrm{ref}})$ removes
all flux-independent features introduced external circuitry
and unintentional reflections.
Multiplying by ${S}_{21,\mathrm{fit}}$ removes the
systematic contribution of the reference,
which would otherwise appear at all values of $\Phi$ as a
static vertically inverted mirror image ($|{S}_{21,\mathrm{fit}}| > 1$)
of the reference resonance.

The scan used as the reference alternates between
$\Phi_{\mathrm{ref}} = \Phi_0/2$ and $\Phi_{\mathrm{ref}} = 0$,
changing from one to the other
whenever $\Phi$ crosses $\Phi_0(1/4 + k/2)$, where $k \in \mathbb{Z}$.
This keeps the resonance in the reference far from the resonance frequency
at the $\Phi$ value being analyzed.
These changes in $\Phi_{\mathrm{ref}}$ cause the apparent discontinuities in
the background color in Figure \ref{data1}.

\section*{Acknowledgements}
We thank Leif Gr\"{o}nberg for depositing the Nb used in this work. We
acknowledge the provision of facilities and technical support by Aalto
University at OtaNano - Micronova Nanofabrication Centre as well as
financial support from the Emil Aaltonen Foundation, the European
Research Council under Grant 278117 (SINGLEOUT), the Academy of
Finland under The COMP Centre of Excellence (251748, 284621) and
grants 257088, 265675, 276528, 286215 and the European Metrology
Research Programme (EMRP EXL03 MICROPHOTON). The EMRP is jointly
funded by the EMRP participating countries within EURAMET and the
European Union. We also acknowldege support from Aalto Centre for
Quantum Engineering.  \bibliography{AEM}
\end{document}